\documentclass[]{emulateapj}
\submitted{Oct. 2007 Astronomical Journal}
\pdfoutput=1
\begin{document}

\title{A Side of Mercury Not Seen by \textit{Mariner 10}}

\author{Gerald Cecil \textup{and} Dmitry Rashkeev}

\affil{Dept. of Physics \& Astronomy, University of North Carolina, Chapel
Hill, NC 27599-3255}

\email{gerald@thececils.org}

\begin{abstract}
More than 60,000 images of Mercury were taken at $\sim29^{\circ}$
elevation during two sunrises, at $\lambda820$ nm, and through a 1.35 m diameter
off-axis aperture on the SOAR telescope. The sharpest resolve
$\sim0\farcs22$ (160 km) and cover $190-300^\circ$ longitude --- a swath unseen
by the \textit{Mariner 10} spacecraft --- at complementary phase angles to previous
ground-based optical imagery. 
Our view is comparable to that of the Moon through weak binoculars.
Evident are the large crater
Mozart shadowed on the terminator, fresh rayed craters, and other 
albedo features keyed to topography and radar reflectivity, including the putative
huge {}``Basin S'' on the limb. Classical bright feature Liguria resolves
across the northwest
boundary of the Caloris basin into a bright splotch centered on a
sharp, 20 km diameter radar crater, and is the brightest feature within
a prominent darker {}``cap'' (Hermean feature Solitudo Phoenicis)
that covers much of the northern hemisphere between longitudes
$140-250^{\circ}$. 
The cap may result from space weathering that darkens via a magnetically enhanced flux of the solar
wind or that reddens low latitudes via high solar insolation.
\end{abstract}


\section{Introduction}

The surface of Mercury is a unique record of early times in our solar
system. But the small angular size of this planet and especially its
proximity in the sky to the Sun limit the clarity of telescopic views.
The \textit{Hubble Space Telescope} can point at Mercury only during {}``twilight''
conditions in Earth shadow, but these observations have never been
attempted. Adaptive optics require for wavefront reference
the use of natural and artificial
guide stars at small airmass whose light would be swamped by
the bright sky. In the mid-1970's, the \textit{Mariner 10} spacecraft made
detailed observations of surface topography ($1-1.5$ km resolution
over a significant area but often at high sun angle), and inferred
magnetic field properties during its mostly successful flybys. Because
of the 3:2 spin-orbit resonance of Mercury, the same hemisphere was
illuminated during all three encounters. 

Over the past 30 yr, the other ``mystery hemisphere'' has been the
target of optical and especially radar imaging to learn if large-scale
morphological structures such as impact basins and their antipodal
effects that were discovered by \textit{Mariner} have counterparts elsewhere.
Radar imagery has covered more than $3/4$ of this hemisphere and has reached 
$\sim5$ km resolution \citep{Harmon07}; sensitive to surface roughness,
tilt, and dielectric constant, radar response does not depend on crater diameter.
A preliminary stratigraphy
of Mercury was developed from the overlap of features in the \textit{Mariner}
images and is keyed to major basin forming events modified by volcanism.
Additional major events recorded on the side not imaged by \textit{Mariner}
could alter this sequence substantially. 

Mercury is a bright object, so exposures on even modest aperture telescopes
can be brief to {}``freeze'' turbulent astronomical seeing. Despite
the large zenith distance and bright sky, selection of the sharpest
{}``lucky images'' from many \citep{Fried78} can permit use of
an aperture large enough to map surface topography. Notable studies
of Mercury using this technique mapped longitudes $270-330^{\circ}$
\citep{Baumgardner00,Dantowitz00} in the morning sky and $210-285^{\circ}$
\citep{Ksanfomality03,Ksanfomality04} \citep[subsequently expanded and summarized in][]{Ksanfomality07}
in the evening. Albedo features on Mercury has also been so imaged
by Webcam equipped amateur astronomers, with impressive results from
modest equipment (F.\ Melillo, private comm.)

To map the lesser explored quadrant $185-300{}^{\circ}$ in morning
illumination (i.e., the complementary phase to that of \citealt{Ksanfomality07}),
in late 2007 March we used a modern high-performance telescope at
an excellent observing site, the 4.1 m SOAR telescope atop Cerro Pachon,
Chile. In \S~2 we describe image acquisition and processing. In \S~3
we present our map of this quadrant, and compare to previous optical
and radar results. In \S~4 we discuss the implications of our findings
on the surface properties of Mercury. We summarize in \S~5.

\section{Methods}

\subsection{Observing Tactics}

\subsubsection{Scheduling}

Ground-level turbulence can be small at dawn after a night of surface
cooling. Our observations were therefore made during a morning elongation
of Mercury that was favorable from Chile and that presented to us
the hemisphere not mapped by \textit{Mariner}. Mercury rotates slowly so
most longitude coverage during an elongation arises from changes in
phase angle. During the second half of the 2007 March-April elongation
Mercury spanned $7\farcs2-6\farcs4$ as its phase increased from 55
to $67\%$ and sub-Earth longitude increased from $178^{\circ}$ to
$222^{\circ}$, while the sub-Earth latitude was $\sim-3.5^{\circ}$.
We intended to observe on 4 mornings every 3 days starting at maximum
elongation, during pre-scheduled University of North Carolina (UNC)
and engineering time.
Unfortunately, weather delivered a sequence of excessive humidity,
clouds, and bad seeing, so we obtained data only on the mornings of
2007 March 23 and April 1. To calibrate, we also observed stars, Jupiter,
and Saturn.

The SOAR telescope is often operated remotely from partner institutions
over the Abeline (Internet2) network \citep{Cecil04}, sharing with
other Cerro-Tololo Inter-American Observatory (CTIO) 
telescopes up to 35 Mbit~s$^{-1}$ bandwidth. Multiple instruments
are installed for the long-term at various SOAR foci.
Our telepresence during dawn and sunrise had minimal impact on other programs.
Because our observations ended after sunrise, certain detector calibrations
that would ideally be done with dawn illumination were instead made
either during evenings or within the light-tight dome.

\subsubsection{Preparing and operating the telescope}

Most modern telescopes mount instruments at Nasmyth foci
where it is particularly challenging to baffle light scatter. Thus,
we did not expect to observe beyond sunrise over the Andes east of
Cerro Pachon. Also, the tertiary and primary telescope mirrors look
upward so are easily contaminated with light scattering dust. Luckily,
the SOAR optics had been washed thoroughly $\sim10$ days before our
observations began.

To obtain sharp images from a stack of tens of thousands of exposures
in seeing characterized by spatial scale $r_{0}$, one must reduce
the telescope aperture $D$ until $D/r_{0}\sim9$ \citep{Fried78}.
$r_{0}$ degrades with zenith angle $z$ and wavelength as $(500\,\mbox{nm}/\lambda)^{1.2}\cos^{-0.6}z,\approx1$
for our study of Mercury. We fabricated and installed across the top
ring of the telescope near the entrance pupil an opaque, black cloth
mask. The mask required an hour to install beginning at the
start of astronomical twilight. Hoping for better than average seeing
at a challenging $\sim27^{\circ}$ elevation, we had our tailor cut
an elliptical hole of minor diameter 1.35 m. This projected a circular
pupil, unobstructed by telescope secondary mirror {}``spider'' supports.

The mask blocked the facility Shack-Hartmann array from sampling the
full set of stellar wavefront tilts to set telescope active optics,
normally done after large-angle motions. The software could not handle
an off-axis sub-aperture. We therefore used only pre-calibrated lookup
tables, indexed exclusively by temperature sensors and the telescope
elevation angle. We set up on a star at $\sim30^{\circ}$ elevation
to confirm focus and pointing, and to make a movie to understand the
current speckle structure. On the first morning, 2007 March 23, long exposure
seeing scaled to the zenith was $\sim0\farcs6$, and through our aperture
we saw what we had hoped to see: a small number of gyrating speckles
of comparable brightness that occasionally merged to produce a very
sharp image. A larger aperture would have passed more speckles, yielding
far fewer coincidences hence sharp images. We could see clear astigmatism
on either side of nominal focus, indicating incomplete setting of
the telescope active optics. However, because our target was centered
and only a few arc-seconds across, astigmatism was useful to maintain
accurate focus so we did not null it. The 2007 April 1 seeing was worse
and we obtained fewer but still excellent images on occasion, probably
because the higher sun angle on the disk enhanced the contrast of
Mercury's subtle shadings.

We acquired Mercury at $15^{\circ}$ elevation, SOAR's limit. These
images were horrible, of course, but improved steadily as sunrise approached.
We recorded occasional crisp detail between $26$ and $30^{\circ}$
elevation before scattered light overwhelmed the signal as the Sun
crested the Andes.

\subsubsection{Camera selection and operation}

We used an Andor Corporation Luca model camera, a thermoelectrically cooled
(stabilized to $-20^{\circ}$C), non-evacuated housing of a Texas
Instruments frame-transfer (electronic shutter) $658\times496$ array
of 10 micron square pixels and $\sim18\%$ QE at $\lambda820$ nm.
The camera records 30 full-frames s$^{-1}$ with nominal 25 electrons
rms readout noise. With a typical acquisition window of $140\times130$
$2\times2$ binned pixels (each $0\farcs06$ on the sky), we acquired
140 frames s$^{-1}$. The camera was connected without reduction optics
directly to a telescope Nasymth focus and to a PC that used Andor's
SOLIS program to acquire data to a SATA drive (49 Mbyte~s$^{-1}$
transfer speed).
A feature of this camera is its electron multiplying gain, adjustable
by software to set the amplification of a separate {}``gain register''
prior to readout. With minimal amplification, exposures of $6.5-8.7$
ms produced peak counts $\sim1/4$ of the camera's 14-bit
digitization range and had negligible readout noise. We used
this setting for all observations and calibrations because we hoped
to work as late as feasible into daylight without saturating the
detector on the brightening sky. To reduce atmospheric dispersion
below one resolution element, we used a $\lambda95$ nm wide interference
filter from CTIO's collection, centered at $\lambda820$ nm, and operating
in the effective f/38 beam.

We took multiple strings of 10,000 exposures stored as $300-400$ Mbyte
FITS format datacubes. A VNC connection from our remote observing
room in Chapel Hill to the computer desktop in Chile allowed us to
control data acquisition and to review representative frames with
ds9. We used the simple \textit{bbcp} program \footnote{See http://www.slac.stanford.edu/~abh/bbcp}
to deliver data expeditiously to UNC at 3 Mbyte~s$^{-1}$
(while maintaining the VNC connection, an audio/video
link to the telescope operator, and telescope/site telemetry). During
each dawn we recorded and transferred $\sim5$ Gbyte of data into
UNC's SOAR archive.

We also recorded stacks of 10,000 dark frames of the same duration
as the data, exposures so short that they showed only fixed pattern
noise near the readout, a 6 DN top-to-bottom gradient in the electronic
bias level, and several tens of {}``hot pixels''. We subtracted
the average dark from each exposure of the datacube.

Although we recorded evening sky flats and morning dome flats, these
unfortunately failed to remove spots in data frames, especially the
March 23 data. Spots seem to arise from contamination on the CCD surface,
not on its dewar window, so are sensitive to details of their specific
illumination. They are noticeable when successive exposures are viewed
as a movie; the dancing planet image is seemingly being viewed through
a somewhat dirty window. Their effect is smeared across the final
stacked image by planet motions between its constituent exposures, but
image contrast on March 23 would have been higher, and confidence
in the reality of subtle albedo variations and shadowing near the
terminator increased, without it. It was also unclear if the nominal
flats calibrated pixel-to-pixel sensitivity variations. To better
match illumination to reduce problems in the flat fields associated
with scattered light from bright dome or sky, we resorted to using
the disks of Jupiter and Saturn. We obtained 1,000 to 10,000 exposure
datacubes of these planets through the filter. The 200 ms
Jupiter exposures showed often exquisite detail in the cloud
belts around, for example, the Little Red Spot, unsurprising given that
this planet was imaged near the zenith in reasonably good seeing.
We coadded the \textit{worst} $10\%$ of the images that were further
blurred by motion between frames, and then used the IRAF \textsf{fit1d}
task to produce spline3 fits row by row and column by column. We combined
row and column fits and ratioed the result into the original image
to obtain a flat field. We scaled the flat to obtain best results
and divided it into each exposure of the datacube.
We retained for further processing the $40,000$ highest elevation
images from the first morning and $20,000$ from the last.

\subsection{Selecting and Processing Sharp Frames}

Key to lucky imaging is how one finds the needle in the haystack.
After exploring and rejecting contrast and wavelet based algorithms,
we simply selected by eye the best image of $\sim500$ from the highest-elevation
string, and then automatically cross-correlated this with the other
tens of thousands to produce an output sequence sorted from narrowest
$50\%$ waist of the two-dimensional correlation peak to fuzziest. Ranking excluded
all double exposures caused by paired clumps of speckles, and most
of the images with fuzzy limbs or gross distortions. Some images deemed
acceptable were in fact distorted ({}``looming''), rotated, or blurred
over part of their area. These were rejected in the next step after
we had loaded the nominally best 500 from the combined stack of all
strings into the \textit{RegiStax} program
(widely used by amateur astronomers to stack Webcam images)
\footnote{See http://www.astronomie.be/registax} to refine the cross-correlation, 
and hence rerank, images within
a $128\times128$ pixel box that spanned the planet.

We selected the final set of images in this sorted sequence of $\sim60$
by eye, drizzle stacking the best 20 and 13 frames from the first
and second mornings, respectively, to form high signal-to-noise ratio final
images with $0\farcs03$ pixels (Fig.\ \ref{fig:SOAR-images-of}, \textit{left images}).
After setting the scale of wavelet number 1 to encompass noise, 
we attenuated its amplitude while boosting 
wavelet scales $2-4$.  The result (Fig. 1, \textit{middle images})
equaled our expectations from a blurred, $0\farcs2$ FWHM version
of a \textit{Mariner} full disk image. As an alternative to wavelets, we
used the smart sharpening filter in Photoshop CS3 followed by $50-100$
iterations of Richardson-Lucy deconvolution (Fig.\ 1, \textit{right images}, \citealt{Richardson72})
to further sharpen bright features but now non-uniquely and with some
added noise. We did not dig out features along the terminator beyond
those immediately apparent (Fig.\ \ref{fig:SOAR-images-of}, \textit{arrows})
because this would have required a much larger final image stack to
attain resolution above the Rayleigh limit \citep{Ksanfomality07}, and because topography
within $15^{\circ}$ of the March 23 terminator had been mapped by \textit{Mariner}.

We drew a circle around the planet image of the correct
radial scale and found its center to $\pm0.5$ pixel. To reduce the
phase-dependent planetary illumination to map albedo variations, we weighted
image intensities by multiplicative factor $[\cos\phi\bullet\max(0.3,\left|\sin\theta\right|)]^{k}$
with $\phi$ the latitude and $\theta$ the longitude difference from
the terminator to the location on the planet and $k=-1/2$ selected
unphysically simply to improve appearances.%

\section{Empirical Results}

Projecting the two enhanced final images made Figure \ref{fig:Radar-topographic-and}$a$,
a cylindrical-equidistant map. Between longitudes $180-300^{\circ}$,
we locate features to better than $2^{\circ}$ near the planetary
equator and to within $3^{\circ}$ near the top of the map, these
uncertainties arising from inexact centers of the half or gibbous
shapes of the planet.%

\begin{figure*}
\begin{center}
    \includegraphics[scale=0.5]{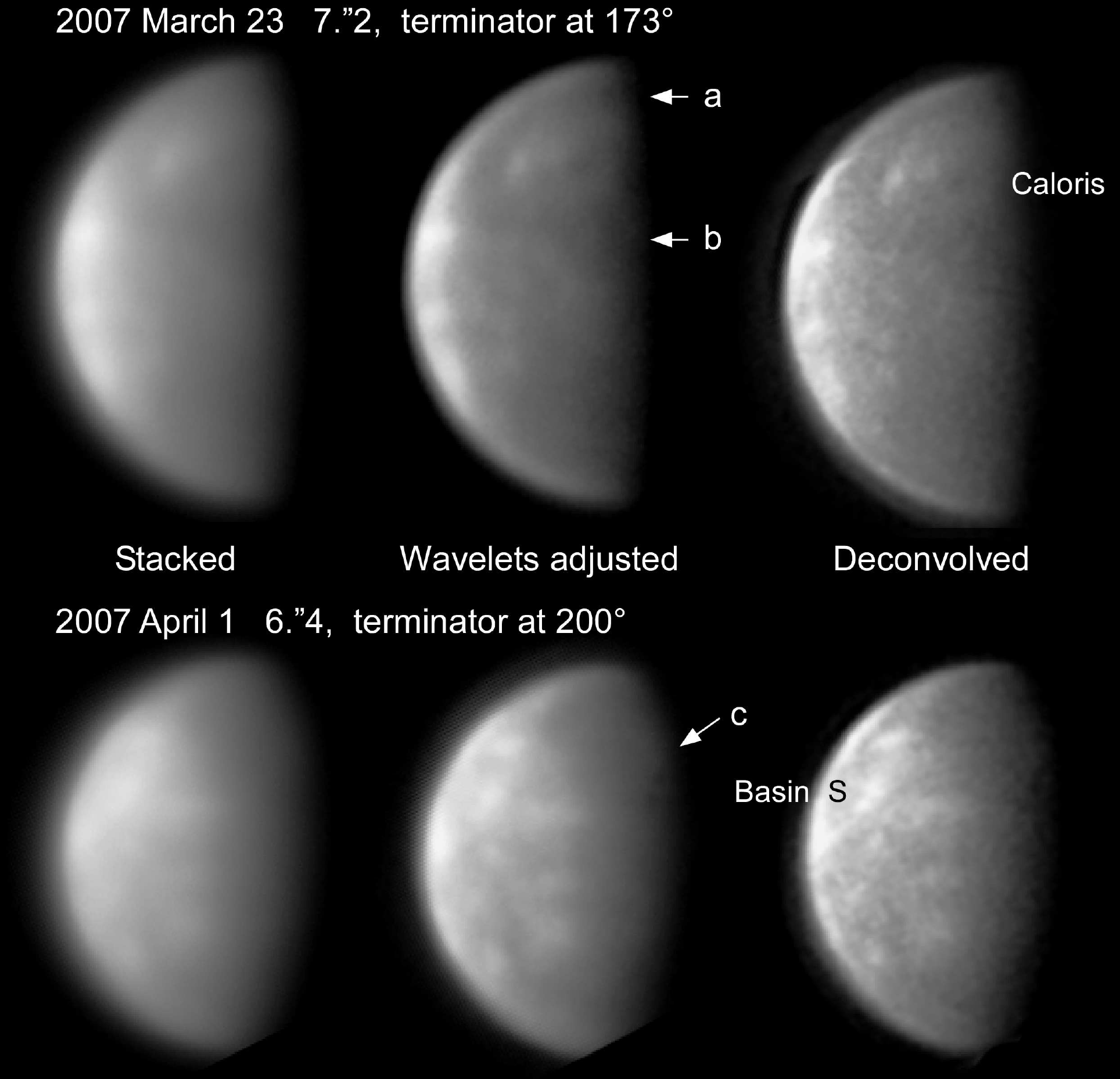}
\caption{\label{fig:SOAR-images-of}{\small SOAR telescope views of a side
of Mercury not imaged by \textit{Mariner 10. Left:} Composites of the sharpest
exposures \textit{(top}, 20 of 40,000 acquired at $29-30^{\circ}$ elevation;
\textit{bottom}, 13 of 20,000 at $25-27^{\circ}$ elevation). They contain
much detail that we have attempted to enhance (\textit{middle}) by boosting
the contribution of intermediate-scale wavelets, or (\textit{right}) by applying
an adaptive sharpening filter followed by $50-100$ iterations of Richardson-Lucy
deconvolution with a plausible PSF. Three isolated bright features
near the left limb in the top row have rotated to disk center in the
bottom where they are revealed in the right column and after
correlation with radar images \citep{Harmon07} to be rayed craters.
Two large basins are labeled on the limbs of the right images:
the Caloris basin and putative {}``Basin S'' proposed by \citet{Ksanfomality07}.
Arrows point to terminator topography discussed briefly in the text.}}
  \end{center}
\end{figure*}

\begin{figure*}
\begin{center}
    \includegraphics[scale=0.87]{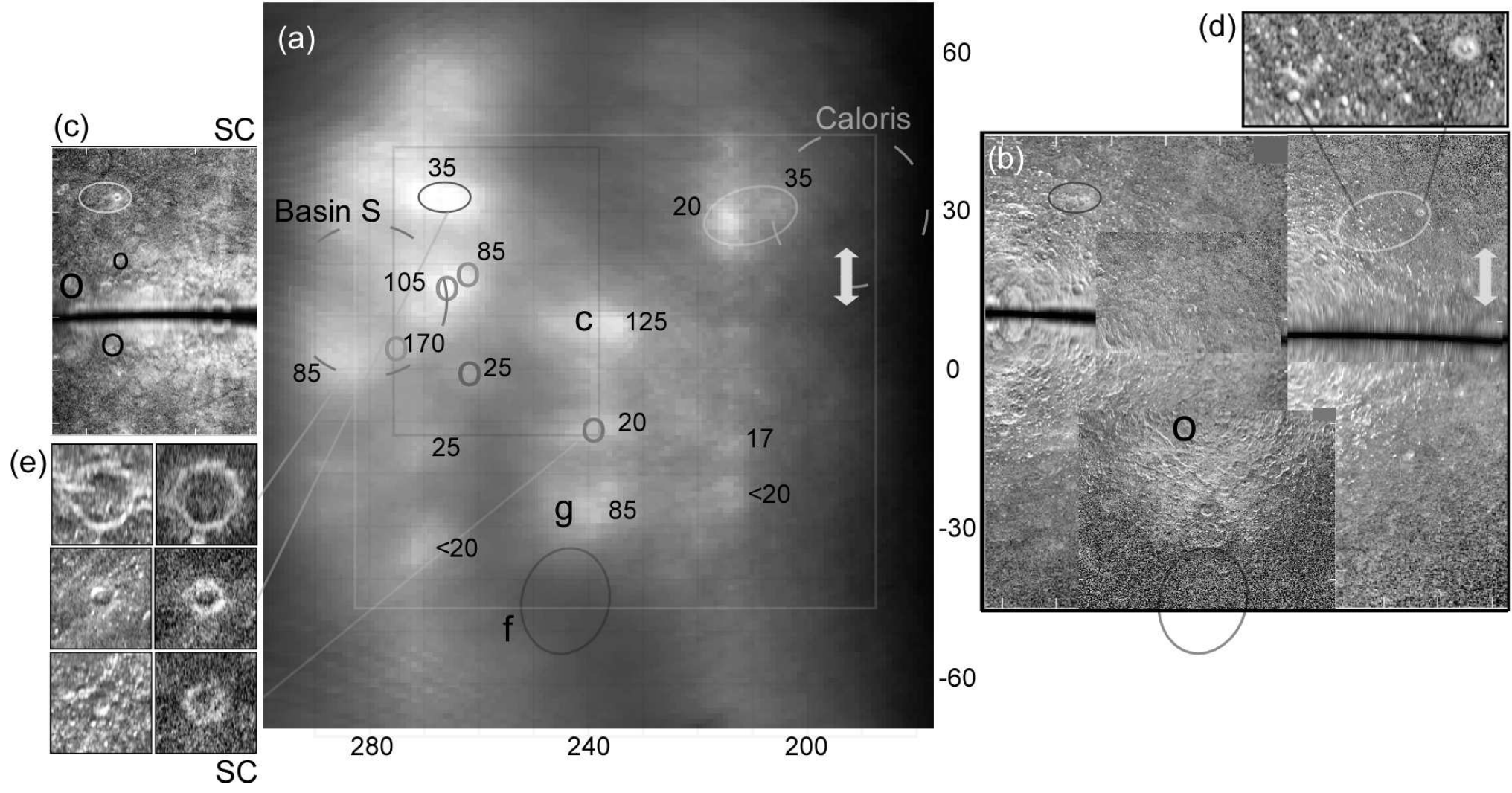}
\caption{{\small \label{fig:Radar-topographic-and}
(a) Map of uncalibrated $\lambda820$
nm albedo variations in cylindrical equidistant projection, constructed
from the two wavelet adjusted SOAR images (Fig.\ \ref{fig:SOAR-images-of}).
The Caloris basin and putative Basin S are denoted by broken circles. 
Bright ramparts of crater Mozart are also evident along the
terminator below Caloris (arrows). Crater diameters (from radar images in \citet{Harmon07}) 
are shown in km.
(b-e) show radar images (Opposite-polarization component unless labeled SC 
for Same component) of craters that coincide
with our bright features. Radar data are ambiguous across the
dark Doppler equators in (b) and (c). 
However, albedo patterns in our data allow us
to identify the circled features of pairs in panel (c) as spurious.}}
\end{center}
\end{figure*}

\begin{figure*}
\begin{center}
    \includegraphics[scale=0.83]{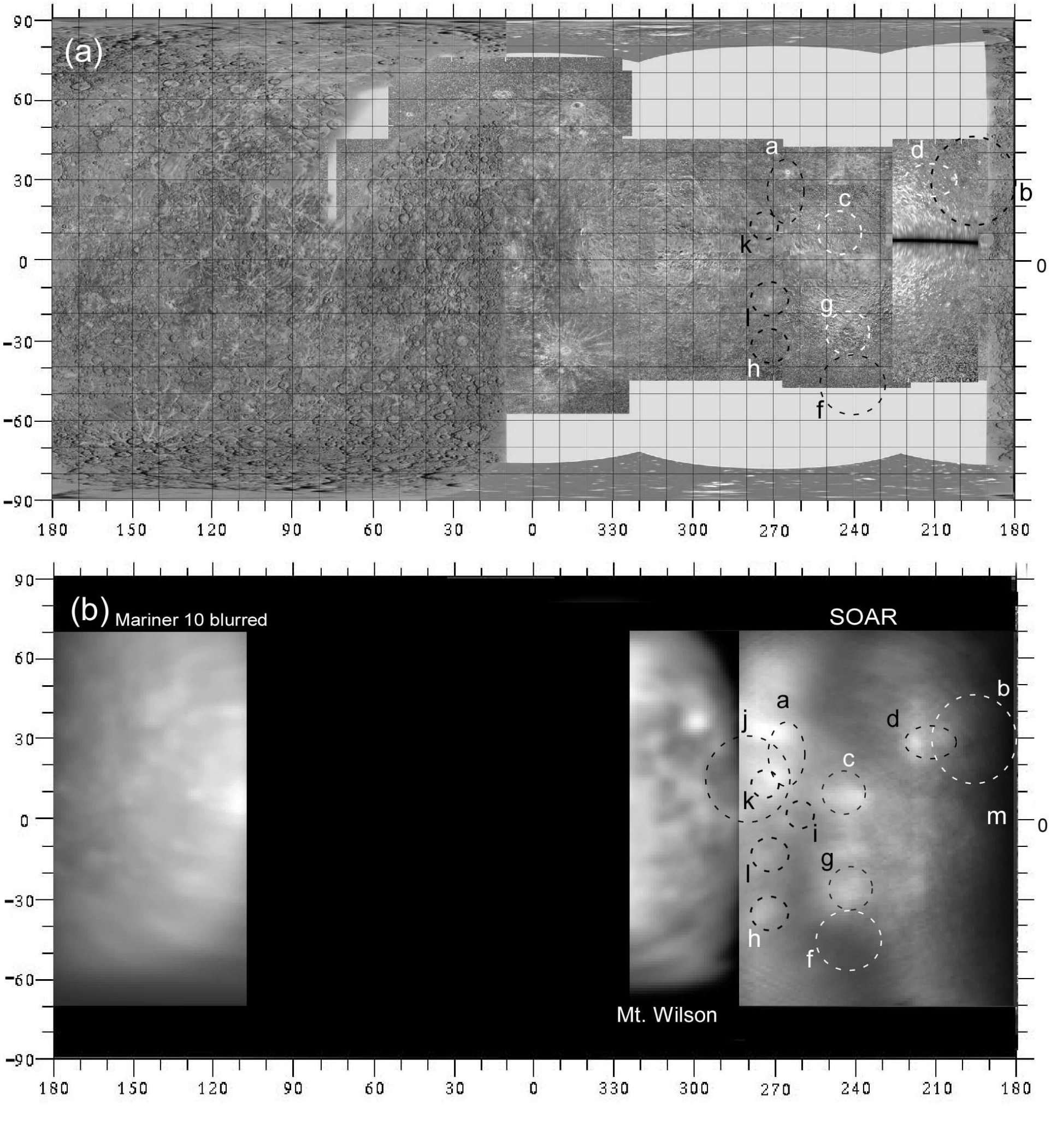}
\caption{\label{fig:Montage-of-most}{\small Montage of some previous maps
of Mercury and our data, cylindrical equidistant projection. (a) \textit{Mariner
10} photo map at left, at right is a mosaic of Arecibo radar images with a few showing
north-south Doppler ambiguities \citep{Harmon07}; (b) A 
\textit{Mariner} mosaic
blurred to our resolution at left, \citet{Ksanfomality04}'s
combination of the \citet{Baumgardner00} and \citet{Dantowitz00}
Mount Wilson images in the middle, and at right our SOAR data. Labeled
features are discussed in the text.}}
\end{center}
\end{figure*}

Our albedo map agrees with the fuzzier but color-calibrated one of 
\citet{Warell01}
from the 0.5 m Swedish Vacuum Telescope used during daylight closer to the zenith.
Figure \ref{fig:Montage-of-most} places our map in the
context of a mosaic of published Arecibo radar images \citep{Harmon07},
and we highlight some agreements below. Correlating images at these
two frequencies is not trivial. As \citet{Harmon07} noted, radar images
whose polarization is opposite to the transmitted beam are sensitive
to sharp surface relief, not shallow gradients, while same-polarization
images respond mostly to wavelength-scale surface roughness and somewhat
to variations of surface dielectric properties. Both image types are
ambiguous across the Doppler equator, but the spurious feature can be rejected by
comparing to other radar scans at different sub-Earth latitude or, as we show below,
to our images.

We tried to compare our map to the crescent image analyzed by \citet{Ksanfomality07},
which overlaps somewhat with our images but has opposite illumination
and foreshortening. What they saw on the bright limb would be evident
in the middle of our April 1 image. The brightest feature in their
image, at $(235^\circ,+32^\circ)$, is undetected in ours. Their bright
feature at $(247^\circ,-7^\circ)$ may be associated with ours
at $(240^\circ,-10^\circ)$. Their crater at $(270^\circ,-15^\circ)$
is our feature {}``l''. Unfortunately, further comparison is unfeasible given the
rapid change in brightness of topographic features as sun angles vary, and the
amplification of the CCD contamination in our March 23 image that would result from
the extensive image processing required to super-resolve details beyond
the telescope-aperture Rayleigh limit.

Figures 2 and 3 reveal albedo variations
tied to radar rayed craters and large dark areas.
For example, dark region {}``j'' in Figure
\ref{fig:Montage-of-most}$b$ centered at $(\sim280^\circ,+16^\circ)$
is the huge {}``Basin S'' posited by \citet{Ksanfomality07}. To
its south and east lie bright radar craters. Feature {}``c'' in
panel Figure \ref{fig:Radar-topographic-and}$a$ is a 125 km diameter crater that \citet{Harmon07}
note as asymmetrically brightened to the north; instead, we see an east-west
extension of its bright material. Bright features {}``a''
in Figure \ref{fig:Montage-of-most} form a broken ring of what radar
image Figure \ref{fig:Radar-topographic-and}$a$ shows are rayed craters
hence must be fresher than an ancient Basin S. These are clearly not
the encircling basin ramparts that \citet{Ksanfomality04} saw shadowed
at lower sun angle. Indeed, \citet{Harmon07} found radar
highlighting only on the western side of the basin, beyond the limb to us.
Bright regions are evident along
the rest of Basin S in the \citet{Baumgardner00} plus \citet{Dantowitz00}
composite Mount Wilson image in the middle of Figure \ref{fig:Montage-of-most}.

Eastward at longitudes $242-250^{\circ}$, prominent bright clumps
at latitudes $+10,-10$, and $-28^{\circ}$ are all radar craters.
In our sharpest images (and Fig.\ \ref{fig:SOAR-images-of} \textit{right images}),
each is a bright core surrounded by a slightly fainter splotch.
In extent, they all are comparable to the largest crater ray systems imaged
at high sun angle by \textit{Mariner} on the opposite hemisphere (Fig.\ \ref{fig:Montage-of-most}$b$
\textit{left}).

Feature {}``g'' in Figure \ref{fig:Radar-topographic-and}$a$ is 200
km across and seems to have radial striations, with Figure \ref{fig:Radar-topographic-and}$c$ showing a
crisp radar crater of diameter 85 km with central peak and a muted
crater more than twice as wide immediately to the south. Further south still
in our images in Figures 2$a$ and 3$b$ is dark region ``f" centered at 
$(240^\circ,-45^\circ)$, Solitudo Persephones in the \citet{Dollfus78} 
IAU albedo map.
Comparable in size to the Caloris basin, {}``f'' is not surrounded
by radar or optical bright features, so is perhaps a plain not a large impact
basin with substantial ramparts. Is it connected physically to {}``g''?

The right (eastern) side of our map spans the \textit{Mariner} unimaged
western boundary of the Caloris basin, {}``b'' in Figure \ref{fig:Montage-of-most}.
Its floor seems to be composed of multiple dark regions. But, intriguingly,
an only slightly brighter but still dark area extends far beyond Caloris,
poleward in a diagonal swath from $(190^\circ,+5^\circ)$ to $(260^\circ,+45^\circ)$;
this is the classical Hermean albedo feature Solitudo Phoenicis in the IAU
map \citep{Dollfus78}.  In fact, despite the lack of photometric calibration, the
\textit{Mariner} mosaic Figure 4 \textit{(right)} shows that this high-latitude darkening
continues eastward to longitude $140^\circ$.

Comparing Figure \ref{fig:Radar-topographic-and}$a$ and $c$, we see that the dark area
sometimes coincides with a change in radar surface texture and has a sharp radar boundary
between $240-250^\circ$.
The dark region seems to be devoid of prominent rayed craters,
implying relatively recent origin, with a striking
exception: bright {}``island d'' in Figure \ref{fig:Montage-of-most}
spans $\sim15{}^{\circ}$ (350 km) at the end of slightly dimmer
IAU Hermean region Liguria
\citep{Dollfus78}.
The closest radar feature in Figure 2$d$ is
a fresh crater at $(203^\circ,+30^\circ)$,
which is, however, several degrees east of our eastern feature.
The optically much brighter western feature at
$(217^\circ,+28^\circ)$ is an inconspicuous 20 km diameter radar crater
that seems to sit on a larger degraded crater (Fig.\ 2d); its
lack of correlation with a prominent radar crater is similar to the situation
for the brightest feature in the Mount Wilson images (Fig.\ 3, \textit{middle}).
On April 1 ``d" lay only $17^{\circ}$ from the terminator, yet was
still so bright that it appears in Figure \ref{fig:SOAR-images-of}
to protrude into the shadowed area (arrow {}``c''). We propose
the name ``Mistral''\footnote{By IAU convention, most surface features on Mercury are named for deceased
artists and writers. Poet Lucila de Maria del Perpetuo Socorro Godoy
Alcayaga (1889-1957, pen name Gabriela Mistral)
was born in Vicu\~{n}a, Chile (visible from the SOAR telescope),
and received the 1945 Nobel Prize in literature.%
} 
for the sharp, optically bright 20 km diameter crater at $(217^\circ,+28^\circ)$.

In our March 23 image, another dark region, Solitudo Atlantis \citep{Dollfus78},
is evident to the south and east, and is bounded to the west by
isolated rayed craters near $210^\circ$. This region is mottled,
indicating its incipient resolution into partially shadowed craters.
Indeed, the illuminated western ramparts of the large crater Mozart
are just detectable (arrow {}``b'') in Figure \ref{fig:SOAR-images-of}.
Finally, dark patch ``a" $(190^\circ,+63^\circ)$ near the terminator 
in Figure 2 is a region without published radar data.

\section{Discussion}

\subsection{{}``Basin S''}

\citet{Ksanfomality04} discovered and argued for this huge ({}``Skinakas'')
basin, centering it at $(280^\circ,+8^\circ)$, $\sim5^{\circ}$ south
of our estimate (which, being at the limb, is less accurate). Both
halves appear near the terminator in the images of \citet{Baumgardner00,Dantowitz00}
and \citet{Ksanfomality03}, respectively, and make it comparable
in size to Caloris. It may be bigger: \citet{Ksanfomality04} showed
that the intensity cut across the middle of this structure is consistent
with two concentric rings of ramparts that extend into the quadrant
that we observed on April 1. Exceeding in extent the lunar south pole-Aitken
feature, if a true impact structure it would be one of the
largest basins in the solar system. The formation of Caloris was the
culminating event in the \textit{Mariner} derived stratigraphy of Mercury.
\citet{Ksanfomality07} assert that Basin S has a degraded appearance,
implying that it is older than Caloris. The ejecta deposits of Basin
S would probably not extend far enough to over/underlap those from
Caloris (permitting direct relative dating from \textit{MESSENGER} spacecraft
imagery during its 2008 gravity-assist flybys of Mercury).

In our April 1 image, Basin S is centered near the bright limb. Within
it, radar shows (Fig.\ \ref{fig:Montage-of-most}$a$) a few intermediate
size craters and indeed we see a bright one, {}``k'', that may be
what \citet{Ksanfomality04} attributed as its central peak (which
would be an impact signature). But the northern half of Basin S is
certainly dark even at high sun angle, and is surrounded on its visible
south and east sides by bright craters, for example, ``a''. In fact, radar
craters account for all the optical bright spots; there are no signs
of boundary ramparts on the E side visible to us.
To the west beyond our limb, \citet{Harmon07}
find radar highlights and speculate that these may be the inner western
rim of Basin S. 
There is no sign in the \textit{Mariner}
imagery of hilly {}``weird terrain'' at the putative Basin S antipode
as is the case for Caloris.

\subsection{\label{sub:Other-dark-features}Other dark features}

As mentioned in \S~3, a dark swath in the 2007 March 23
image extends westward at reduced prominence in the April 1 image and eastward in the \textit{Mariner}
mosaic (Fig.\ 4).
The boundary between lower and higher albedo regions is sharp, as
is the boundary of the {}``island'' of bright features {}``d''.
The radar image Figure \ref{fig:Radar-topographic-and}$c$ also shows a sharp boundary,
but only between $240-250^\circ$ does it coincide with the albedo change.
The darker {}``cap'' is a striking asymmetry across Mercury;
there are indications of a counterpart darkening at high southern latitudes
in Figure \ref{fig:Montage-of-most}$b$. 
Is it superficial, a result of space weathering that darkens and/or reddens
an exposed surface? We have only monochrome red images, so cannot
map colors. One way to darken the surface is an enhanced influx of
charged particles near the magnetic poles \citep{Killen01}, although Mercury's
field is supposed to be strong enough to keep
the solar wind from reaching the surface most of the time \citep{Russell88}. 
Alternatively, equatorial and
lower latitude regions could be reddened (hence brightened in our
filter) relative to the poles by weathering from enhanced solar irradiation
\citep{Hapke01}.
If the darkening is instead due to a fundamental change in surface properties 
(as suggested by the change in radar texture), this asymmetry would
cause Mercury to resemble the other terrestrial planets.

\begin{figure*}
\begin{center}
    \includegraphics[scale=0.48]{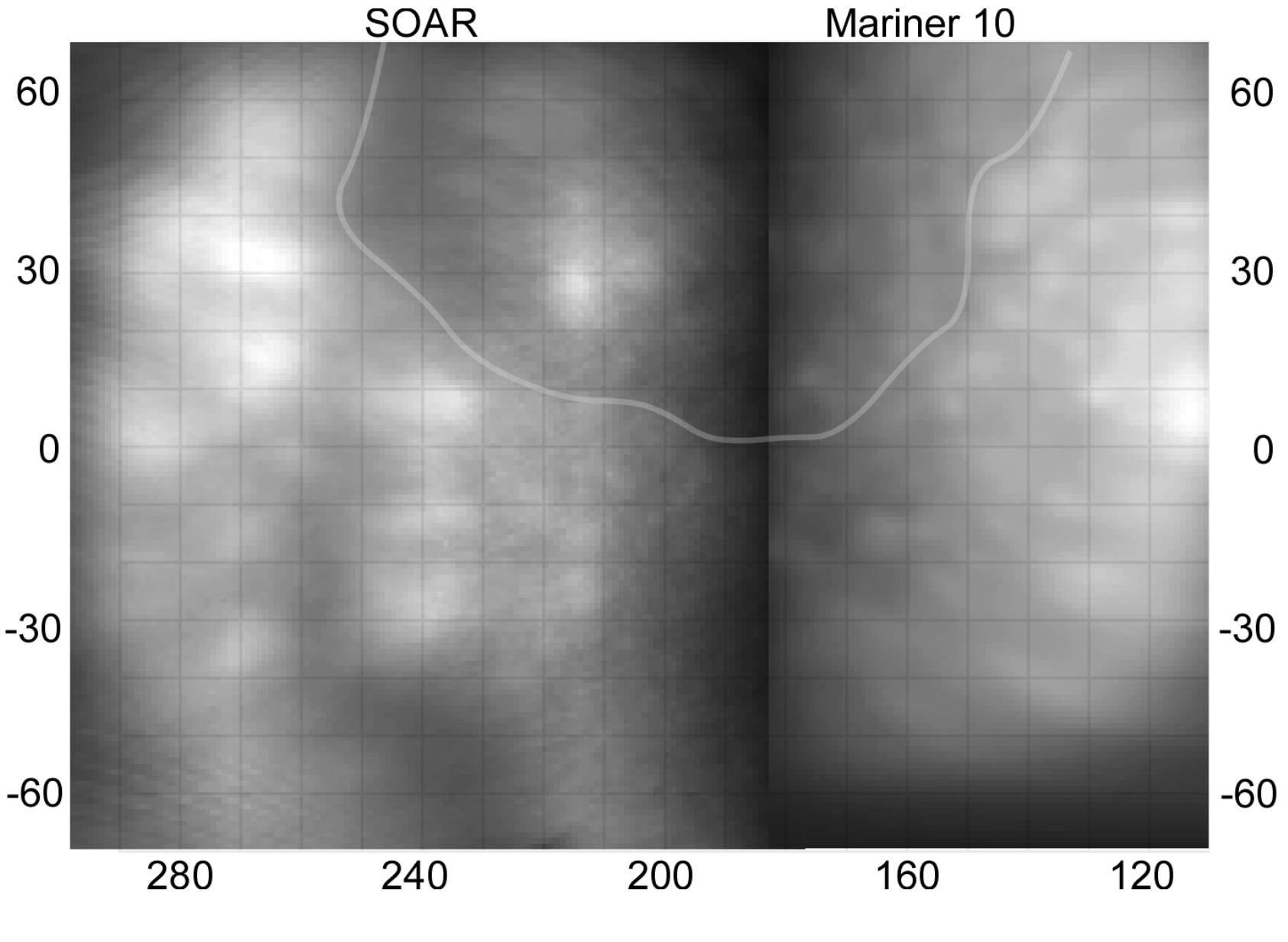}
\caption{{\small \label{fig:Mariner-10-departure}
\textit{Right:} Blurred \textit{Mariner 10} mosaic image; \textit{left:}
SOAR map, where the $\cos\theta$ (with $\theta$ the angle to
the sun from the vertical at each point on the planet's surface) illumination pattern has been
removed. Outlined at top is the approximate extent of the darker region
that extends poleward from northern mid-latitudes over this hemisphere.}}
\end{center}
\end{figure*}

\subsection{Volcanoes?}

How volcanism may have shaped the planet's surface is controversial.
No definitive volcanic feature was identified even in \textit{Mariner 10} 
close-ups.
A non-Lunar aspect is the extensive distribution of inter-crater plains,
attributed to either volcanism or obscuring basin ejecta. \citet{Robinson97}
attempted to calibrate the compromised \textit{Mariner 10} two-filter photometry,
and have compared the resulting color variations to lunar measurements.
In their view, changes in surface color are due both to space weathering
and to intrinsic compositional variations consistent with volcanic
fire fountains.

Long-term volcanism is a possibility following the radar measurements
by \citet{Margot07} of variations in the planet's forced longitudinal
libration that are consistent with mantle slippage inertia over a
partially molten core. Such a core structure also explains the weak
magnetic field discovered by \textit{Mariner}. (Alternatively, \citealt{Ksanfomality07}
noted that \textit{Mariner 10} passed over Basin S, and posit that
the measured field is a relic of this putative impact.) The absence
of global tectonics other than the overall contraction inferred from
planet-wide scarps \citep{Strom75} may permit substantial volcanic
structures to accumulate slowly over time, not just as fire fountains
in the distant past. Shield volcanoes could be flat enough to be subtle
in existing radar images that are sensitive only to short-wavelength
tilts; radar altimetry \citep{Clark88} has not spanned large areas of the planet. 
Nevertheless, all bright features in our map can be attributed to
rays or ramparts of radar craters.

\section{Summary}

Our ``lucky image" stacks bettered $0\farcs25$ resolution and have mapped
prominent rayed craters and other features across part of the hemisphere
unseen by \textit{Mariner 10} but subsequently imaged by radar. The region
complements that studied by \citet{Ksanfomality07} also using short
exposure stacks. They and \citet{Ksanfomality04} have posited
a large dark feature ({}``Basin S'' or {}``Skinakas basin'')
near $(280^\circ,+10^\circ)$. Although we observed it near the limb,
we do see its darker floor even at the high sun angle that brightens
some adjacent craters. However, we find only bright radar craters around it, and
no features that could be attributed to basin 
ramparts on its eastern boundary.

Several topographic features are observed at the terminator, including
ramparts of the large crater Mozart. A very bright spot that is a sharp 
radar crater lies within a dark swath
that extends across at least $140-250^{\circ}$ longitude at latitudes
reaching northward of $30^\circ$ at $180^{\circ}$ longitude to
northward of $+45^{\circ}$ latitude at $250^{\circ}$. No other prominent
rayed craters, even at higher sun angles, lie within this region.
Starting in 2011, \textit{MESSENGER}'s intensive mapping of compositional variations
and topography as it orbits will bring Mercury's history into ever
sharpening focus, and hopefully will require no luck whatsoever.

\acknowledgements{}

We thank SOAR personnel in Chile, specifically Eduardo Serrano for
overseeing design and tailoring of the fabric pupil mask, observer
supporters Patricio Ugarte, Daniel Maturana, Alberto Alvarez, and
Sergio Pizarro for extended shifts into the early morning, and SOAR
director Steve Heathcote for his enthusiastic support. We thank Andor
Corporation for loan of a Luca camera to debug our procedures and Andrei Tokovinin
for use of his camera to take the data.
John Harmon provided helpful interpretation of his radar images.
Figure 3a is based on work by Philip Stooke.
This project used only personal funds, but would not have
succeeded without the remote observing facility at UNC, which was
equipped by private donations to the Dept. of Physics and Astronomy.

\end{document}